\RequirePackage[british]{babel}
\documentclass[reqno,a4paper,12pt]{article}
\usepackage[latin1]{inputenc}
\usepackage[a4paper,hmargin=2cm,tmargin=3cm,bmargin=3cm]{geometry}
\usepackage{amsmath,amssymb,amstext,amsthm,amscd}
\usepackage{mathrsfs,eucal,graphicx}
\usepackage{color}
\usepackage{verbatim}
\usepackage{array}
\usepackage[nosort]{cite}
\usepackage[vcentermath]{youngtab}
\Yboxdim{4pt}
\usepackage{hyperref}
\theoremstyle{plain}

\theoremstyle{definition}

\allowdisplaybreaks
\begin{document}
\renewcommand{\theequation}{\arabic{section}.\arabic{equation}}
\title{\bf $D$=$3$ (p,q)-Poincar\'e supergravities from Lie algebra
expansions \bigskip \bigskip}

\author{J.A. de Azc\'{a}rraga, \\
Dept. Theor. Phys. and IFIC (CSIC-UVEG), \\
\medskip
\medskip
Univ. of Valencia, 46100-Burjassot (Valencia), Spain\\
 J. M Izquierdo,\\
Dept. Theor. Phys., Univ. of Valladolid, \\
47011-Valladolid, Spain}
\bigskip
\bigskip
\date{{\small  July 13, 2011}}
\maketitle
\begin{abstract}
We use the expansion of superalgebras procedure (summarized in the
text) to derive Chern-Simons (CS) actions for the ($p,q$)-Poincar\'e
supergravities in three-dimensional spacetimes.
After deriving the action for the  $(p,0)$-Poincar\'e supergravity as
a CS theory for the expansion $osp(p|2;\mathbb{R})(2,1)$ of
$osp(p|2;\mathbb{R})$, we find the general $(p,q)$-Poincar\'e superalgebras
and their associated $D$=3 supergravity actions as CS gauge theories from
an expansion of the simple $osp(p+q|2,\mathbb{R})$ superalgebras,
namely $osp(p+q|2,\mathbb{R})(2,1,2)$.
\end{abstract}
\newpage

\section{Introduction and results}
Some important limits in physics can be described in terms of
Lie algebra contractions \cite{Seg:51,IW:53,Sal:61,W-W:00 }.
For instance, the simple de Sitter $so(4,1)$ and anti-de Sitter
$so(3,2)$ algebras lead by contraction to the $D$=4 Poincar\'e algebra.
The contraction parameter may be related to the AdS$_4$ constant curvature when
$SO(3,2)$ is interpreted as the isometry group
of four-dimensional spacetime (it is $1/R$ where $R$ is the
radius of the universe), or to the square root of the
cosmological constant $\Lambda$ when the algebra is taken as the starting
point for the construction of gravity via gauging (for
$so(4,1)$ the cosmological constant changes sign).
The most familiar example -which in fact motivated the idea \cite{Seg:51,IW:53}-
is the Galilei algebra as a $c$$\rightarrow$$\infty$ \.In\"on\"u-Wigner (I-W)
contraction of the Poincar\'e one. Of course, the procedure
also applies to superalgebras: for instance, the (${\mathcal N}$=1) $D=4$
superPoincar\'e algebra is an I-W contraction of the
fourteen-dimensional $osp(1|4)$ superalgebra, the even part of
which is the $adS_4$ algebra $so(3,2)\sim sp(4,\mathbb{R})$.

Contractions can take Lie algebras that are direct sums of two Lie
algebras into others that no longer have a direct sum structure
or that in general alter the structural relation
among the original algebras. The oldest example is the centrally
extended eleven-dimensional Galilei algebra, which is
a $c$$\rightarrow$$\infty$ I-W contraction of the direct sum of
the Poincar\'e algebra and $u(1)$. This quantum-mechanical
Galilei algebra arises because the $c$$\rightarrow$$\infty$ limit
involves both terms of the direct sum \cite{Sal:61, AA:85};
the supersymmetric case was discussed in \cite{dAz-Gin:91}.
Another example, in $D$=2, is the centrally extended
Poincar\'e algebra which is used to construct the
$D$=1+1 CGHS model \cite{CGHS:92} of gravity as a gauge
theory. As noted in \cite{Can-Jack:92}, the relevant centrally (magnetic-like)
extended $D$=1+1  Poincar\'e algebra, of dimension four,
may be obtained by a  contraction. Such a contraction was
called `unconventional' in \cite{Can-Jack:92} but it is, in fact,
an ordinary I-W contraction involving
the direct sum of the three-dimensional $adS_2=so(1,2)$ and a
one-dimensional algebra. As with the contraction of the trivially
extended $D$=4 Poincar\'e algebra to the centrally extended Galilei one,
this I-W contraction involves both algebras
in the direct sum\footnote{The name `unconventional contraction' is
often used to denote {\it standard}
IW contractions where the generators of the original algebra are rescaled
at the same time that new, {\it additional} generators are introduced.
As a result, the contraction being performed is in fact an ordinary
I-W contraction of a {\it larger} algebra, not of the original one.
The re-scaling of the generators is singular in the contraction limit,
but allows for the contraction in the algebra commutators (see \cite{AA:85,CUP} for
the cohomological meaning of the procedure and \cite{Sal:61} for the earliest
work).}, and the dimensions (3+1) of the original  and
contacted algebras are the same (the term
`extension' is used throughout in its mathematical sense).

  The supersymmetric version of the CGHS model was shown in
\cite{Rive:93} to be the gauge theory of an algebra obtained by a
similar trick, which in this case corresponds to performing an I-W contraction
of a semidirect extension of the five-dimensional super de Sitter algebra
$osp(1|2;\mathbb{R})$ (which is the supersymmetric version of $so(1,2)\,$)
by a four-dimensional abelian algebra which contains two
bosonic charges and a pair of fermionic generators.
The resulting (5+4)-dimensional contracted algebra  is the superalgebra suitable
for two-dimensional dilaton supergravity \cite{Rive:93},
rather than the five-dimensional ${\mathcal N}$=1 $D$=1+1 superPoincar\'e algebra
which is obtained by an I-W contraction of $osp(1|2;\mathbb{R})$
by rescaling the two bosonic `translations' and
the two fermionic generators, respectively, by $\mu$
and $\mu^{1/2}$. In fact, the superalgebra in \cite{Rive:93}
is an extension of the ${\mathcal N}$=1, $D$=2 superPoincar\'e
one by the four-dimensional ideal of the additional generators,
in which one of the bosonic generators is central. For other examples of this type
of I-W contractions (in planar physics) see \cite{Hor-Ply-Va:10}.

The anti-de Sitter algebra for $D$-dimensional spacetime is
$so(D-1,2)$. For $D$=3, it is $adS_3=so(2,2)=so(1,2)\oplus so(1,2)$,
which is not simple; its supersymetric generalization is
$osp(1|2;\mathbb{R}) \oplus osp(1|2;\mathbb{R})$. There are
actually two non-isomorphic $osp(1|2;\mathbb{R})$ algebras,
$osp(1|2;\mathbb{R})_{\pm}$ (see Sec.~\ref{ospdet} for the
$\pm$ signs). In \cite{Ach-Tow:86} a set of $D$=3 AdS-type supergravities was
given as Chern-Simons (CS) models based on the superalgebras
$osp(p|2;\mathbb{R})_+ \oplus osp(q|2;\mathbb{R})_- \equiv adS(p,q)$ in
the terminology of \cite{Ach-Tow:89}. A problem appeared when taking
the Poincar\'e limit of these $AdS(p,q)$ CS supergravities for $p\geq 2$ or
$q\geq 2$ (the action does not have a limit). The only consistent
Poincar\'e limit yields \cite{Ach-Tow:89} the Marcus-Schwarz $D=3$
${\mathcal N}$-extended Poincar\'e theories \cite{Mar-Sch:83},
for which the superalgebra does not include the gauge $so(p)$ and $so(q)$
generators. The difficulty was overcome for a class of
($p,q$)-Poincar\'e supergravities, related to the $AdS(p,q)$ anti-de Sitter ones
(the (1,0) case being the old three-dimensional example in \cite{1001:83}).
Since the problem to obtain a Poincar\'e limit of the $AdS(p,q)$ CS theories was due to the
$so(p)$ and $so(q)$ gauge fields terms in the CS action in \cite{Ach-Tow:86}, the authors
of \cite{Ho-Iz-Pa-To:95} started from a larger algebra, the trivial extension
$osp(p|2;\mathbb{R})_+ \oplus osp(q|2;\mathbb{R})_- \oplus so(p) \oplus so(q)$
of $adS(p,q)$ by $so(p) \oplus so(q)$. This superalgebra was used
to take a particular I-W contraction that gave a well defined
Poincar\'e limit, here denoted $s\mathcal{P}(p,q)$, to obtain
the associated ($p,q$)-Poincar\'e $\mathcal{N}$-supergravities,
$\mathcal{N}=p+q$. Clearly\footnote{dim$\,osp(m|q;\mathbb{R})$=
${1\over 2}m(m-1)$+${1\over 2}q(q+1)+mq\,,\,q=2n$, where $mq$ is the
dimension of the odd part; the even subalgebra is $so(m)\oplus sp(q)$.
For $q$=2, dim$\,osp(m|2;\mathbb{R})\,$=$\,{1\over 2}m(m-1)+3+2m$. },
\begin{equation}
\label{Ppq}
\mathrm{dim}\,s\mathcal{P}(p,q)= p(p-1)+q(q-1)+6+2(p+q),
\end{equation}
in which $2(p+q) = 2{\mathcal N}$ is the dimension of the odd sector.
These $D$=3 ($p,q$)-Poincar\'e supergravity  models \cite{Ho-Iz-Pa-To:95}
constitute the subject of this paper\footnote{It may be worth mentioning that,
in another context, the analysis of (2+1)-dimensional gravity
with higher-spin fields (or HS AdS gravity) has also led to the appearance of larger
algebras, namely ${\mathcal W}$-algebras \cite{Hen-Rey:10,Cam-Fre-Pfe-The:10}.}.

  We provide here an alternative derivation of both the $s\mathcal{P}(p,q)$
superalgebras and of the associated $D$=3 ($p,q$)-Poincar\'e CS supergravities
by means of a construction directly based on the $osp({\mathcal N}|2;\mathbb{R})$
superalgebra. Rather than extending trivially the non-simple $adS(p,q)$ superalgebra to
then perform a I-W contraction, we will use a comparatively novel procedure,
the {\it expansion of algebras}, which leads to new algebras and superalgebras by
`expanding' the original ones. This technique was used first in  \cite{Hat-Sak:01}
and studied in general in \cite{AzIzPiVa:02,AzIzPiVa:04}, where
it was called the {\it expansion method} (see also \cite{Iza-Ro-Sal:06}
for further developments). By series expanding the Maurer Cartan forms
in the MC equations of the original algebra and then looking at the
equations that result by equating equal powers in the expansion
parameter, it is  possible to retain a number of the coefficient
one-forms in the different series in such a way that the resulting
equations become the MC ones of a new Lie (super)algebra,
the expanded one. Clearly, the dimension of the
expanded algebra is {\it larger} in general
than that of the original algebra. It may be seen
\cite{AzIzPiVa:02,AzIzPiVa:04}, nevertheless, that the expansion
procedure also includes the general Weimar-Woods
contractions \cite{W-W:00} (which in turn include the I-W ones)
as a particular case, for which the expansion is of
course dimension-preserving. In \cite{Hat-Sak:01,AzIzPiVa:02} various sets
of conditions were provided to cut the series expansions of the MC forms
in such a way that the retained one-form coefficients
satisfy the MC equations of a new, expanded
Lie (super)algebra, the structure constants of which
follow from those of the original algebra.

The expansion procedure is convenient to generate new, larger
(super)algebras from a given one\footnote{For instance, the
algebra in \cite{Rive:93} already discussed may be shown to be given by a
central extension of the expansion $osp(1|2;\mathbb{R})(2,3,2)$
by a one-dimensional algebra; dimensionally (see eq.~\eqref{dimsuper}
below) one can check that [2.1+2.2+1.2]+1=9.}. Besides, it
presents computational advantages, particularly for
model building. For instance, starting from a CS model based on a simple
Lie algebra and then expanding the gauge field one-forms exactly
in the same way as the MC forms, every power in the expanded CS model provides
\cite{AzIzPiVa:02} a CS action for a certain expanded Lie algebra.
This fact was used  to obtain \cite{AzIzPiVa:02} the $D=3$
${\mathcal N}$=1 Poincar\'e supergravity \cite{1001:83} action as a CS
gauge theory by expanding an $osp(1|2;\mathbb{R})$-based CS model in a
parameter $\lambda$ ($[\lambda]=L^{-\frac{1}{2}}$) and by looking at the
term in $\lambda^2$. This $D$=1+2 Poincar\'e superalgebra or
$s\mathcal{P}(1,0)$ (which is the expansion $osp(1|2;\mathbb{R})(2,1)$
\cite{AzIzPiVa:02}) can also be obtained as an I-W contraction of
the eight-dimensional direct sum $osp(1|2;\mathbb{R}) \oplus sp(2)$,
but the expansion method leads directly to the $s\mathcal{P}(1,0)$
superalgebra and to the CS action of the (1,0)-Poincar\'e
supergravity.

It is clear that one cannot expect that every contraction
of an extended group be given by an expansion, since the
expansion procedure has much less freedom than the combination
of extensions (which involve pairs of algebras) and
contractions. So it makes sense to investigate whether the $s\mathcal{P}(p,q)$
superalgebra and the $(p,q)$-Poincar\'e supergravities can be obtained
from expansions of a certain CS model based on a simple superalgebra when
$p+q > 1$, as we already know to be the case for $(p,q)$=$(1,0)$ (or
$(0,1)$). We shall show that the  $D=3$ ($p,q$)-superPoincar\'e algebra
$s\mathcal{P}(p,q)$, obtained in \cite{Ho-Iz-Pa-To:95}
as an I-W contraction of $adS(p,q)\oplus (so(p)\oplus so(q))$
which mixes the $so(p)$ and $so(q)$ generators present in both
parts of the direct sum (and that cannot be obtained
by contracting a simple superalgebra), can be derived
from the expansion $osp(p+q|2;\mathbb{R})(2,1,2)$ of $osp(p+q|2;\mathbb{R})$.
The process will involve taking the quotient of the expansion by an
ideal. The gauge fields associated with the generators of this
ideal will not appear in the expanded CS lagrangian, which
coincides with that of the $D=3$ ($p,q$)-Poincar\'e
CS supergravity models given in \cite{Ho-Iz-Pa-To:95}.
This is the result of this paper: the gauge superalgebra
$s\mathcal{P}(q,p)$ underlying the $D$=2+1 ($p,q$)-Poincar\'e
supergravities as well as their corresponding CS actions can be obtained
from the expansion  $osp(p|2;\mathbb{R})(2,1)$ when $q$ (say) is zero
and, when  $p,q\not=0$, from $osp(p+q|2;\mathbb{R})(2,1,2)$, in this
case by virtue of the physical irrelevance in the action of a bosonic
ideal of this superalgebra.

   The plan of the paper is as follows. In Sec.~2 we recall the
properties of $osp(p|2;\mathbb{R})$ trough its MC equations to
fix the notation and write the corresponding gauge free differential
algebra (FDA); similarly, Sec.~3 reviews the ingredients of the expansion
procedure relevant here. Sec.~\ref{p0sugra} considers the case
$(p,0)$-Poincar\'e supergravity, where it is shown that the expanded
algebra $osp(p|2;\mathbb{R})(2,1)$ is the $(p,0)$-superPoincar\'e one,
$s\mathcal{P}(p,0)$. In Sec.~\ref{pqgral} the general $(p,q)$ case is examined,
and in particular the ($p,q$)-Poincar\'e supergravity CS lagrangian is obtained
from the expansion $osp(p+q|2;\mathbb{R})(2,1,2)$. Sec.~5 contains some final
remarks.

\section{The superalgebra $osp(p|2;\mathbb{R})$}
\label{ospdet}
\setcounter{equation}0

$OSp(p|2;\mathbb{R})$ is the supergroup of
transformations preserving the $(p+2)$$\times$$(p+2)$
orthosymplectic metric $C$, $g^t C g =C$,
where $g$ is a $(p,2)$-type supermatrix, $t$ indicates
supertranspose, and $C$ is given by
\begin{equation}
\label{metric}
       C=\left( \begin{array}{cc}
         \epsilon_{\alpha \beta} & 0 \\
         0 &  -iI_{p\times p}\\
       \end{array} \right)
\end{equation}
in which $\epsilon_{\alpha\beta}$ is the $2\times 2$ symplectic metric.
Therefore, the superalgebra elements $X\in osp(p|2;\mathbb{R})$ satisfy
$X^t C+C X=0$. Let then $\mathfrak{a}$ be a $osp(p|2;\mathbb{R})$-valued MC one-form.
The previous condition means that $\mathfrak{a}$  will have the
block form structure
\begin{equation}
\label{MCforms}
       \mathfrak{a}=\left( \begin{array}{cc}
         l^\alpha{}_\beta & i\nu^{j \alpha} \\
         \nu^i_\beta & h^{ij} \\
       \end{array} \right) \quad , \quad
       \alpha, \beta=1,2 \; ; \; i,j,=1,\dots,p \quad,
\end{equation}
or one where the $i$ at the top right box is removed and
a $-i$ is added to the bottom left box. In \eqref{MCforms},
$\nu^i_\alpha=\epsilon_{\alpha\beta}\nu^{\beta i}$,
$(\nu^{\alpha i})^*=\nu^{\alpha i}$, $h^{ij}=-h^{ji}$ (the
euclidean indices $i,j$ will be always up), and the symplectic
subalgebra matrix $l^\alpha{}_\beta$ has the property
$l_{\alpha\beta} = l_{\beta\alpha}$.

    The MC equations for the supermatrix form \eqref{MCforms} read
\begin{equation}
\label{MCeqs}
     d\mathfrak{a} =- \mathfrak{a}\wedge \mathfrak{a} \qquad i.e. ,
\end{equation}
\begin{equation}
\label{MCeqs1}
   \left(%
\begin{array}{cc}
 dl^\alpha{}_\beta   & id\nu^{j\alpha} \\
         d\nu^i_\beta & dh^{ij}\\
\end{array}%
\right) = - \left(%
\begin{array}{cc}
l^\alpha{}_\gamma& i \nu^{k \alpha} \\
\nu^i_\gamma & h^{ik}    \\
\end{array}%
\right) \wedge
\left(%
\begin{array}{cc}
  l^\gamma{}_\beta & i \nu^{j \gamma} \\
         \nu^k_\beta &  h^{kj}\\
\end{array}
\right) \ .
\end{equation}
Explicitly, the $osp(p|2;\mathbb{R})$ MC equations are
\begin{eqnarray}
\label{MCcomp}
  dh^{ij} &=& - h^{ik}\wedge h^{kj}- i\nu^{i}_\gamma
  \wedge \nu^{j\gamma} \ ,
  \nonumber\\
  d\nu^{j\alpha} &=& -  h^{jk}\wedge \nu^{k\alpha} -
  l^\alpha{}_\gamma \wedge \nu^{j\gamma} \ ,
  \nonumber\\
  dl^\alpha{}_\beta &=& -l^\alpha{}_\gamma \wedge
  l^\gamma{}_\beta -i\nu^{k\alpha} \wedge \nu^{k}_\beta\ .
\end{eqnarray}
For the other possible choice of the matrices $\mathfrak{a}$, the
MC equations have a plus sign in the two fermion bilinears
above. These two $osp$ algebras, inequivalent as real algebras
and denoted $osp(p|2;\mathbb{R})_+$ and $osp(p|2;\mathbb{R})_-$
respectively, correspond to taking a different sign for the
$\{Q_\alpha,Q_\beta\}$ anticommutator \cite{Ach-Tow:89};
clearly, their complexified versions coincide. Unless otherwise stated,
$osp(p|2;\mathbb{R})$ will mean $osp(p|2;\mathbb{R})_+$ as given
by eq.~\eqref{MCcomp}.

The curvatures $\mathbb{F}$ of the gauge
fields $\mathbb{A}$ associated with the different
$osp(p|2;\mathbb{R})$ generators, needed for constructing a CS model,
are now directly obtained from the MC equation \eqref{MCeqs}
by writing
\begin{equation}
\label{gaugecur}
    \mathbb{F} = d\mathbb{A} + \mathbb{A} \wedge \mathbb{A}
\end{equation}
in terms of the `soft' forms $\mathbb{A}$, which satisfy
eqs.~(\ref{MCcomp}) when the curvatures vanish. Writing the matrix
blocks of $\mathbb{A}$ and $\mathbb{F}$
respectively as
\begin{equation}
\label{AFmatrix}
     \mathbb{A} = \left(%
\begin{array}{cc}
  f^\alpha{}_\beta & i\xi^{j \alpha} \\
  \xi^j_\beta &  A^{ij}\\
\end{array}%
\right) \ , \quad
 \mathbb{F} = \left(%
\begin{array}{cc}
 \Omega^\alpha{}_\beta&  i \Psi^{j \alpha} \\
  \Psi^i_\beta &  F^{ij} \ \\
\end{array}
\right) \; ;
\end{equation}
eq.~\eqref{gaugecur} gives
\begin{eqnarray}
\label{curvatures}
 F^{ij} &=& dA^{ij} + A^{ik} \wedge A^{kj} + i \xi^{i}_\gamma \wedge \xi^{j\gamma} \ ,
 \nonumber \\
 \Psi^{j\alpha} &=& d\xi^{j\alpha} + f^\alpha{}_\gamma \wedge \xi^{j\gamma}
  + A^{jk}\wedge \xi^{k\alpha} \ , \nonumber\\
 \Omega^\alpha{}_\beta &=& df^\alpha{}_\beta + f^\alpha{}_\gamma \wedge
 f^\gamma{}_\beta + i \xi^{k\alpha} \wedge \xi^{k}_\beta \ .
 \end{eqnarray}

   Similarly, the infinitesimal gauge variations characterized by
 a matrix $\phi$
\begin{equation}
\label{Bmatrix}
   \phi = \left(%
\begin{array}{cc}
 \tilde{b}^{\alpha}{}_\beta & i\tilde{\zeta}^{j \alpha} \\
  \tilde{\zeta}^i_\beta & \tilde{a}^{ij} \\
\end{array}%
\right)
\end{equation}
and are given by
\begin{equation}
\label{gaugevar}
  \delta_\phi \mathbb{A} = d \phi + \mathbb{A} \phi -
  \phi \mathbb{A} \ ,\quad
\delta_\phi \mathbb{F} = \mathbb{F}
  \phi- \phi \mathbb{F}.
\end{equation}
In components, the above equations read
\begin{eqnarray}
\label{exptr}
\delta\phi A^{ij} &=& d \tilde{a}^{ij} + A^{ik} \tilde{a}^{kj} - \tilde{a}^{ik} A^{kj} +
i\xi^i{}_\gamma \tilde{\zeta}^{j\gamma} - i\tilde{\zeta}^i{}_\gamma \xi^{j\gamma} \; ,
\nonumber\\
\delta_\phi \xi^{j\alpha} &=& d \tilde{\zeta}^{j\alpha} + \tilde{a}^{kj} \xi^{k\alpha} +
f^\alpha{}_\gamma \tilde{\zeta}^{j\gamma} - A^{kj} \tilde{\zeta}^{k\alpha}
- \tilde{b}^\alpha{}_\gamma \xi^{j\gamma} \; ,
\nonumber\\
\delta_\phi f^\alpha{}_\beta &=& d \tilde{b}^\alpha{}_\beta + f^\alpha{}_\gamma
\tilde{b}^\gamma{}_\beta - \tilde{b}^\alpha{}_\gamma f^\gamma{}_\beta +
i\xi^{k\alpha} \tilde{\zeta}^k{}_\beta - i\tilde{\zeta}^{k\alpha} \xi^k{}_\beta \; ,
\nonumber\\
\delta_\phi F^{ij} &=& F^{ik} \tilde{a}^{kj} - \tilde{a}^{ik} F^{kj} +
i\Psi^i{}_\gamma \tilde{\zeta}^{j\gamma} -i \tilde{\zeta}^i{}_\gamma \Psi^{j\gamma} \; ,
\nonumber\\
\delta_\phi \Psi^{j\alpha} &=& \tilde{a}^{kj} \Psi^{k\alpha} +
\Omega^\alpha{}_\gamma \tilde{\zeta}^{j\gamma} - F^{kj} \tilde{\zeta}^{k\alpha}
- \tilde{b}^\alpha{}_\gamma \Psi^{j\gamma} \; ,
\nonumber\\
\delta_\phi \Omega^\alpha{}_\beta &=& \Omega^\alpha{}_\gamma
\tilde{b}^\gamma{}_\beta - \tilde{b}^\alpha{}_\gamma \Omega^\gamma{}_\beta +
i\Psi^{k\alpha} \tilde{\zeta}^k{}_\beta - i\tilde{\zeta}^{k\alpha} \Psi^k{}_\beta \; .
\end{eqnarray}

\section{Chern-Simons actions from the expansion procedure}
\label{exp+CS}
\setcounter{equation}0

   Expansions of a Lie algebra $\mathcal{G}$ are obtained by series
expanding some of its MC forms $\omega^i$, $i=1,\dots,\mathrm{dim}\,\mathcal{G}$,
in terms of a parameter $\lambda$. These series
follow by re-scaling by a power of $\lambda$ some group coordinates
$g^i$ in the expression of the MC forms as given, say, by
the left invariant one-forms on the associated group manifold (see \cite{AzIzPiVa:02}).
When the expansions of the MC forms are introduced in
the MC equations, an infinite-dimensional algebra is formally
obtained by requiring that each power in $\lambda$ vanishes separately.

Finite Lie algebras are obtained by cutting the expansions of the MC
one-forms in such a way that the coefficient one-forms
$\omega^{i,\alpha}$ accompanying $\lambda^\alpha$ in
$\omega^i(g,\lambda)=\sum_{\alpha=0}^\infty\lambda^\alpha\omega^{i,\alpha}$
satisfy equations (obtained by equating equal powers in $\lambda$)
that become the MC equations of a new algebra, the
{\it expanded algebra}. Various sets of conditions that guarantee that
it is possible to retain consistently a finite number of $\omega^{i,\alpha}$ satisfying
the MC equations of the new expanded (super)algebra were given
in \cite{Hat-Sak:01, AzIzPiVa:02}. The result that is relevant in
this paper is the following.

Let $\mathcal{G}= V_0 \oplus V_1 \oplus V_2$ be a Lie
superalgebra, $V_1$ its Grassmann odd part, $V_0 \oplus V_2$ the
even one, and $V_0$ a subalgebra of $\mathcal{G}$.
Let $\omega^{i_0}$, $\omega^{i_1}$ and $\omega^{i_2}$ the MC forms dual to the algebra basis
generators in the vector subspaces  $V_0$, $V_1$ and $V_2 \subset \mathcal{G}$
respectively; obviously, $i_0=1,\dots , \mathrm{dim} V_0$, $i_1=1,\dots ,
\mathrm{dim} V_1$, $i_2=1,\dots , \mathrm{dim} V_2$. Let the
MC equations of $\mathcal{G}$ be written in the form
\begin{equation}
\label{MCsup} d \omega^{k_s} = - \frac{1}{2} c^{k_s}_{i_p j_q}
\omega^{i_p} \wedge
 \omega^{j_q} \qquad p,q,s= 0,1,2 \; .
\end{equation}
Then, it is consistent to expand the MC forms of $V_0\oplus V_2$ in
terms of even powers of $\lambda$ and those of $V_1$ in terms of
odd powers of $\lambda$, up to the orders below, as
(see \cite{{AzIzPiVa:02}}, Sec.~5)
\begin{eqnarray}
\label{expth}
   \omega^{i_0} &=& \sum^{N_0}_{\alpha_0=0,\ \alpha_0 \, even}
\lambda^{\alpha_0} \omega^{i_0,\alpha_0} \;, \nonumber \\
 \omega^{i_1} &=& \sum^{N_1}_{\alpha_1=1,\ \alpha_1 \, odd}
\lambda^{\alpha_1} \omega^{i_1,\alpha_1} \; ,\nonumber\\
 \omega^{i_2} &=& \sum^{N_2}_{\alpha_2=2,\ \alpha_2 \, even}
\lambda^{\alpha_2} \omega^{i_2,\alpha_2} \; ,
\end{eqnarray}
{\it i.e.}, we may keep the one-form coefficients up to
$\omega^{i_1,N_1}, \,\omega^{i_2,N_2},\, \omega^{i_3,N_3}$,
provided that the even $N_0$, $N_2$ and
odd $N_1$ integers satisfy one of the three conditions
below
\begin{eqnarray}
\label{conditions}
  & & N_0= N_1+1 =N_2 \; ; \nonumber\\
 & & N_0= N_1-1 = N_2 \; ; \nonumber\\
 & & N_0=N_1-1=N_2-2 \; .
\end{eqnarray}
This means that the $\omega^{i_1,\alpha_1},\omega^{i_2,\alpha_2},\omega^{i_3,\alpha_3}$
one-forms, where the $i_s=1,...$dim$V_s \,(s=1,2,3)$ and the $\alpha_s$
have the ranges in the sums above, define formally the MC forms of finite-dimensional
superalgebras. These {\it expanded superalgebras}, subordinated to the
splitting of the vector space of the original superalgebra
$\mathcal{G}=V_0\oplus V_1\oplus V_2$, will be denoted by
$\mathcal{G}(N_0,N_1,N_2)$. Their dimension is given by
\begin{equation}
\label{dimsuper}
\textrm{dim}\,\mathcal{G}(N_0,N_1,N_2)=
\left(\frac{N_0+2}{2}\right)\textrm{dim}V_0 +
\left(\frac{N_1+1}{2}\right)\textrm{dim}V_1 +
\left(\frac{N_2}{2}\right)\textrm{dim}V_2 \quad ,
\end{equation}
and are determined by the following structure constants
\begin{equation}
\label{expsc}
   C^{k_s,\alpha_s}_{i_p,\beta_p\, j_q,\gamma_q} =
\left\{\begin{array}{cc}
  0 & \; \mathrm{if} \beta_p+\gamma_q \neq \alpha_s \\
 c^{k_s}_{i_p j_q} & \; \mathrm{if} \beta_p+\gamma_q = \alpha_s\\
\end{array}\right.  \; ,
\end{equation}
which are expressed in terms of those of the original algebra $\mathcal{G}$.
As stated, the above result is obtained by substituting
\eqref{expth} in \eqref{MCsup}, identifying equal powers of
$\lambda$, and taking $N_0$, $N_1$ and $N_2$ in such a way that
the resulting MC equations describe a superalgebra.

   From the above expanded Lie superalgebras it is possible to obtain,
by moving to the corresponding generic `soft' forms,
$\omega^{i_s,\alpha_s} \rightarrow A^{i_s,\alpha_s}$, their gauge
free differential algebra (FDA) and gauge variations (of infinitesimal
parameters $\varphi^{i_s,\alpha_s}$)
\begin{eqnarray}
\label{expFDA}
 F^{k_s,\alpha_s} &=& dA^{k_s,\alpha_s}+ \frac{1}{2}
C^{k_s,\alpha_s}_{i_p,\beta_p\, j_q,\gamma_q} A^{i_p,\beta_p}
\wedge
A^{j_q,\gamma_q} \ , \nonumber\\
d F^{k_s,\alpha_s} &=&  C^{k_s,\alpha_s}_{i_p,\beta_p\,
j_q,\gamma_q} F^{i_p,\beta_p} \wedge
A^{j_q,\gamma_q} \ , \nonumber\\
\delta A^{k_s,\alpha_s} &=& d \varphi^{k_s,\alpha_s} -
C^{k_s,\alpha_s}_{i_p,\beta_p\, j_q,\gamma_q}
\varphi^{i_p,\beta_p} \wedge A^{j_q,\gamma_q} \ .
\end{eqnarray}
One important point, which will be used in what follows, is that
eqs.~\eqref{expFDA} may also be obtained by inserting in the
equations of the gauge FDA associated with $\mathcal{G}$,
\begin{eqnarray}
\label{FDA} F^{k_s} &=& dA^{k_s} + \frac{1}{2} c^{k_s}_{i_p j_q}
A^{i_p} \wedge
A^{j_q} \ , \nonumber\\
d F^{k_s} &=& c^{k_s}_{i_p j_q} F^{i_p} \wedge
A^{j_q} \ , \nonumber\\
\delta A ^{k_s} &=& d \varphi^{k_s}- c^{k_s}_{i_p j_q}
\varphi^{i_p} \wedge A^{j_q} \ ,
\end{eqnarray}
the formal expansions of $A^{k_s}$, $F^{k_s}$ and $\varphi^{k_s}$
in terms of  $A^{k_s,\alpha_s}$, $F^{k_s,\alpha_s}$ and
$\varphi^{k_s,\alpha_s}$ with exactly the same structure as
that of the MC forms $\omega^{k_s}$ in \eqref{expth},
and then identifying equal powers of $\lambda$.
\medskip

\noindent
{\it Expansions and CS models}

Lie algebra expansions may be used to obtain, from a given
gauge Chern-Simons (CS) model based on a Lie (super)algebra
$\mathcal{G}$, new CS ones associated with the different expansions
of $\mathcal{G}$ (see \cite{AzIzPiVa:02,AzIzPiVa:04} for details).
Let $\mathcal{G}$ be a Lie superalgebra, for instance of the type described
above, and let $k_{i_1,\dots i_l}$ be the coordinates of a symmetric invariant
$l$-tensor of $\mathcal{G}$, where $i=1,\dots,\mathrm{dim}\,\mathcal{G}$.
Then, it follows that the $2l$-form $H$ constructed out of the
curvatures $F$,
\begin{equation}
\label{2lform}
   H= k_{i_1,\dots i_l} F^{i_1} \wedge \dots \wedge F^{i_l} \quad ,
\end{equation}
is closed and gauge invariant and may be
taken as the starting point in a Chern-Weil construction. Since
gauge FDAs are contractible \cite{Su:77,Ni:83}, they have
trivial de Rham cohomology so that $H$
defines a $(2l-1)$-form (the CS form) $B$, such that
$dB=H$. Then, integrating $B$ over a
$(2l-1)$-dimensional manifold $\mathcal{M}^{2l-1}$, a CS
model is obtained through the action
\begin{equation}
\label{CSaction}
  I[A] = \int_{\mathcal{M}^{2l-1}} B(A) \; .
\end{equation}

  It is possible now to perform an expansion of the $A^i$ and $F^i$
fields of the form \eqref{expth} ({\it i.e.},
$A^i=\sum_{\alpha_p} \lambda^{\alpha_p} A^{i_p,\alpha_p}$
and similarly for $F^i$). This leads to an expansion of
the action \eqref{CSaction} given by
\begin{equation}
\label{CSexpansion}
 I[A,\lambda] = \int_{\mathcal{M}^{2l-1}} B(A,\lambda) =
 \int_{\mathcal{M}^{2l-1}} \sum_{N=0}^\infty \lambda^N B_N(A) =
\sum_{N=0}^\infty \lambda^N I_N[A] \ ,
\end{equation}
where in the sums $N$ is even.
The same expansion, when applied to
\eqref{2lform}, leads to
\begin{equation}
\label{Hexp}
  H(F,\lambda) = \sum_{N=0}^\infty \lambda^N H_N \ ,
\end{equation}
where, necessarily,
\begin{equation}
\label{CSHexp}
   H_N = d B_N(A) \; .
\end{equation}
This means that, for each $N$, the actions
\begin{equation}
\label{NCS}
  I_N = \int_{\mathcal{M}^{2l-1}} B_N(A)
\end{equation}
define CS models obtained by the expansion of the original,
$\mathcal{G}$-based one. Given $N$,
the Lie algebra corresponding to the CS action $I_N$ is the one
that contains all the gauge curvatures (field strengths)
included in $H_N$. The most
common situation is that all the curvatures
$F^{k_0,\alpha_0}$ (for $\alpha_0= 0, 2, \dots N$),
$F^{k_2,\alpha_2}$ (for $\alpha_2= 2, 4, \dots N$) and
$F^{k_1,\alpha_1}$ (for $\alpha_1= 1, 3, \dots N-1$) are present
in $H_N$, so that the Lie superalgebra corresponding to $I_N$ is
$\mathcal{G}(N,N-1,N)$, which is the first case in
\eqref{conditions}. It may be, however, that not all the field
strengths actually appear in the CS action \eqref{NCS}. This
will be the case for the expansion of $osp(p+q|2;\mathbb{R})$
of interest here, as will be shown at the end of Sec.~\ref{pqgral}.

\section{The $(p,q)$-Poincar\'e supergravities from
an expansion of $osp(p+q|2;\mathbb{R})$}
\setcounter{equation}0

Consider first a natural CS model associated with
$osp(p|2;\mathbb{R})$. It is constructed from the closed,
invariant four form
\begin{equation}
\label{simpleH}
   H= \textrm{Tr}(\mathbb{F}\wedge \mathbb{F}) \; ,
\end{equation}
where $\mathbb{F}$ is given in
eq.~\eqref{AFmatrix}, $\textrm{Tr}$ stands for the supertrace and
we have moved back to the notation of Sec.~\ref{ospdet}. $H$ is
exact, $H=dB$, and the potential CS form $B$ is gauge invariant
up to the exterior derivative of a two-form. As a result, the CS action
functional
\begin{equation}
\label{simpleI}
  I[A]=\int_{\mathcal{M}^3} B =  \int_{\mathcal{M}^3}
\textrm{Tr}\left( \mathbb{F}\wedge \mathbb{A} - \frac{1}{3}
\mathbb{A}\wedge \mathbb{A} \wedge \mathbb{A} \right)
\end{equation}
is gauge invariant up to boundary terms that depend on the
topology of $\mathcal{M}^3$ and on the type of gauge
transformations considered (`large' {\it vs.} `small'). Since the field
equations are not affected by these boundary terms, they are gauge
invariant, and it is easy to check that they are given by the
vanishing of the curvatures, $ \mathbb{F}= 0$.

Using the $osp(p|2;\mathbb{R})$ components of the $\mathbb{F}$ and
$\mathbb{A}$ matrices given in \eqref{AFmatrix}, the above four-form
$H$ and CS action $I[A]$ turn out to be
\begin{eqnarray}
\label{CSsimcom}
  H &=& \Omega^\alpha{}_\gamma \wedge \Omega^\gamma{}_\alpha
- F^{ik} \wedge F^{ki} - 2i\Psi^i_\gamma \wedge \Psi^{i\gamma} \quad ,
\nonumber \\
 I[A] &=& \int_{\mathcal{M}^3} \Omega^\alpha{}_\gamma \wedge
f^\gamma{}_\alpha - \frac{1}{3} f^\alpha{}_\gamma \wedge
f^\gamma{}_\delta \wedge f^\delta{}_\alpha - f^\alpha{}_\gamma
\wedge \xi^{l\gamma} \wedge \xi^l_\alpha
\nonumber\\
 & & - F^{ik}\wedge A^{ki} + \frac{1}{3} A^{ik} \wedge
A^{kl} \wedge A^{li} - i \xi^{k\alpha} \wedge A^{kl} \wedge
\xi^l_\alpha \nonumber \\
& & +2i \Psi^{k\alpha}\xi^k_\alpha \quad , \quad i,l,k=1,\dots, p\quad .
\end{eqnarray}

Taking the above $OSp(q|2;\mathbb{R})_+$ CS action minus a copy
of that corresponding to $OSp(q|2;\mathbb{R})_-$ with indices
$i',k'=1,\dots,q$, and identifying appropriately the gauge fields,
the CS actions of the $AdS(p,q)$ supergravities \cite{Ach-Tow:86,Ach-Tow:89}
are obtained. As already mentioned, the limit of
these $AdS(p,q)$ supergravities when the AdS dimensionful parameter is
taken  to zero is not straightforward. A way out
was found in \cite{Ho-Iz-Pa-To:95} using a CS model based on a larger
superalgebra, $osp(p|2;\mathbb{R})_+ \oplus osp(q|2;\mathbb{R})_-\oplus so(p) \oplus so(q)$.
The I-W contraction limit was then taken after making suitable
linear combinations of the $so(p)$ generators in
$osp(p|2;\mathbb{R})_+\oplus so(p)$ and $so(q)$ in $osp(q|2;\mathbb{R})_-\oplus so(q)$.
In this way, the I-W contraction limit gives the $s\mathcal{P}(p,q)$
superalgebra, the gauge $so(p)\oplus so(q)$ fields remain
coupled to the gravity fields, and the action
of the $(p,q)$-Poincar\'e supergravities is obtained. Clearly,
$adS(p,q) \nsim adS(q,p)$ but, as we shall see,
$s\mathcal{P}(p,q)\sim s\mathcal{P}(q,p)$. It is not
difficult to see why this is so: the difference in sign
for the  $\{Q_\alpha,Q_\beta\}$ anticommutators
in $adS$ cannot be absorbed by redefining the generators in their
$r.h.s.$, something that can be done for superalgebras of
Poincar\'e type.

We  derive now the $D$=3 $(p,q)$-Poincar\'e
supergravities by means of the expansion procedure. Since the cases $(p,0)$
(or $(0,q)$) and $(p,q)$, $p\neq 0$, $q\neq 0$, turn out to be
slightly different, they will be treated separately.

\subsection{$D$=$3$ $(p,0)$-Poincar\'e supergravity as a
CS theory of the expansion $osp(p|2;\mathbb{R})(2,1)$}
\label{p0sugra}
${}$\\
We begin with a $osp(p|2;\mathbb{R})$-based model\footnote{It is not
difficult to check that the same ($p,0$)-Poincar\'e supergravity would
follow from $osp(p|2;\mathbb{R})_-\;$, since the sign differences
in the $osp(p|2;\mathbb{R})_{\pm}\,$ algebras are
reabsorbed and disappear at the Poincar\'e level.} and
perform an expansion of the $\mathbb{F}$ and $\mathbb{A}$ fields
in the action, eq.~\eqref{simpleI} or \eqref{CSsimcom}.
It is possible to assign physical dimensions to the expansion parameter
$\lambda$ so that the coefficient one-forms in the
expansion are dimensionful too (the original $osp(p|2;\mathbb{R})$ gauge
fields are dimensionless, as it has to be the case for a simple
algebra with dimensionless structure constants). In order to obtain
gravity actions, it is convenient to take $[\lambda]=L^{-\frac{1}{2}}$ in
geometrized units. The next
step is to identify the vector spaces $V_0$, $V_1$ and $V_2$ of
Sec. 3. The space $V_1$ has to be associated with the odd
generators, so its dual is generated by the fermionic one-form fields
$\xi^{i\alpha}$. Here it is convenient to choose $V_2=0$ and $V_0$
as the whole $so(p)\oplus sp(2)\subset osp(p|2;\mathbb{R})$
bosonic subalgebra, to which the bosonic $A^{ij}$ and
$f^\alpha{}_\beta$ one-form fields are associated.

The expansion $osp(p|2;\mathbb{R})(2,1)$ is determined by that
of the MC forms in \eqref{MCcomp} up to the
orders $N_0=2$ (bosonic part) and $N_1=1$ (fermionic part).
This means that the gauge forms and field strengths associated with
$osp(p|2;\mathbb{R})(2,1)$ are those that correspond to
$osp(p|2,\mathbb{R})$, eqs.~\eqref{AFmatrix}, expanded
up to these orders, namely,
\begin{eqnarray}
\label{pexpan}
 f^\alpha{}_\beta & =&  \omega^\alpha{}_\beta + \lambda^2
e^\alpha{}_\beta + o(\lambda^4)\nonumber \\
A^{ij} &=& A^{ij,0} + \lambda^2 A^{ij,2} + o(\lambda^4)\nonumber \\
\xi^{i\alpha} &=& \lambda \psi^{i\alpha} + o(\lambda^3) \quad ;
\nonumber\\
 \Omega^\alpha{}_\beta & =&  R^\alpha{}_\beta + \lambda^2
T^\alpha{}_\beta + o(\lambda^4)\nonumber \\
F^{ij} &=& F^{ij,0} + \lambda^2 F^{ij,2} + o(\lambda^4)\nonumber \\
\Psi^{i\alpha} &=& \lambda \mathcal{D}\psi^{i\alpha} + o(\lambda^3) \quad,
\quad i,j=1,\dots, p\;,
\end{eqnarray}
where we have put
$f^{\alpha \beta,0}\equiv \omega^{\alpha\beta},
f^{\alpha \beta,2}\equiv e^{\alpha\beta},
\xi^{i\alpha,0} \equiv \psi^{i\alpha}$ and
$\Omega^{\alpha\beta,0}\equiv R^{\alpha\beta}$ and
$\mathcal{D}\psi^{i\alpha}$ is defined in the next formula.
The relations among the $osp(p|2;\mathbb{R})(2,1)$ gauge fields and
field strengths above may be found from the general expressions
in eqs.~\eqref{expFDA} and \eqref{expsc}, or directly by expanding
eqs.~\eqref{curvatures}. They are given by
\begin{eqnarray}
\label{pcurv}
R^\alpha{}_\beta &=& d \omega^\alpha{}_\beta +
\omega^\alpha{}_\gamma \wedge \omega^\gamma{}_\beta \; ,
\nonumber\\
T^\alpha{}_\beta &=&  de^\alpha{}_\beta +
\omega^\alpha{}_\gamma \wedge e^\gamma{}_\beta +
e^\alpha{}_\gamma \wedge \omega^\gamma{}_\beta + i\psi^{k\alpha} \wedge \psi^k_\beta
\nonumber \\
 & & \equiv De^\alpha{}_\beta + i\psi^{k\alpha} \wedge \psi^k_\beta \; ,
 \nonumber \\
  F^{ij,0} &=& dA^{ij,0} + A^{ik,0} \wedge A^{kj,0} \ , \nonumber\\
F^{ij,2} &=& dA^{ij,2} + A^{ik,0}\wedge A^{kj,2} + A^{ik,2}\wedge A^{kj,0} +
   i\psi^i_\gamma \wedge \psi^{j\gamma} \ ,\nonumber\\
\mathcal{D}\psi^{j\alpha} &=& d \psi^{j\alpha} +
\omega^\alpha{}_\gamma \wedge \psi^{j\gamma} + A^{jk,0}\wedge
\psi^{k\alpha} \equiv D \psi^{j\alpha} +
A^{jk,0}\wedge \psi^{k\alpha} \; .\\ \nonumber
\end{eqnarray}

In view of the above, it will be natural to introduce $D$=3 imaginary gamma matrices
and make the following identifications\footnote{In terms of Pauli matrices,
$\gamma^0=\sigma^2$, $\gamma^1= i\sigma^1$, $\gamma^2=i\sigma^3$
(mostly minus metric); the antisymmetrized product $\gamma^{ab}$
is taken with weight one.}:  $\omega^\alpha{}_\beta \propto \omega_{ab}
(\gamma^{ab})^\alpha{}_\beta$, which is dimensionless $[\omega^\alpha{}_\beta]=L^0$,
is the Lorentz spin connection in three spacetime dimensions;
$R^\alpha{}_\beta \propto R_{ab} (\gamma^{ab})^\alpha{}_\beta$ is its curvature;
$e^\alpha{}_\beta \propto  e_a (\gamma^a)^\alpha{}_\beta$,
$[e^\alpha{}_\beta]= L^1$, may be identified with the dreibein and
its curvature, $T^\alpha{}_\beta \propto T_a (\gamma^a)^\alpha{}_\beta$, with torsion;
$\psi$ and $\mathcal{D}\psi$, with length
dimensions $[\psi]=[\mathcal{D}\psi]= L^{\frac{1}{2}}$, are the
gravitino one-form field and its complete (Lorentz plus gauge)
covariant derivative; $A^{ij,0}$ are the dimensionless
gauge one-form fields, $[A^{ij,0}]= L^0$, and $F^{ij,0}$ their
curvatures. Finally, the $SO(p)$ one- and two-forms
$A^{ij,2},\, F^{ij,2}$ have length dimensions $L^1$.

The above dreibein, spin connection and gravitino identifications are
justified by the form of the action \eqref{NCS} for $N=2$ that follows
from the expansion of \eqref{simpleI}, which becomes
the supergravity action. The choice  $N=2$ is selected because then the
action $I_2$ has length dimensions $[I_2]= L^1$, which are the
dimensions of a $D$=3 action in geometrized units ($L^{D-2}$ for an action
on general $D$-dimensional spacetimes). Indeed, $I_2$ is given by
\begin{equation}
\label{actionP0}
    I_2 = \int_{\mathcal{M}^3} (2 R^\alpha{}_\gamma
    \wedge e^\gamma{}_\alpha -2 F^{ik,0} \wedge A^{ki,2}
    -2i \mathcal{D} \psi^i_\gamma\wedge \psi^{i\gamma}) \ ,
\end{equation}
which can be seen to coincide with the $N=2$ term of the expansion of
the second line in \eqref{CSsimcom} except for the integral of an
exact form on $\mathcal{M}^3$. The integrand in \eqref{actionP0}
is a potential form of the $\lambda^2$ term in the expansion of
the first line of \eqref{CSsimcom} as in \eqref{Hexp}, which is
given by
\begin{equation}
\label{HP0}
    H_2 = 2 R^\alpha{}_\gamma
    \wedge T^\gamma{}_\alpha- 2 F^{ik,0} \wedge F^{ki,2} -
    2i \mathcal{D} \psi^i_\gamma \wedge \mathcal{D} \psi^{i\gamma} \; .
\end{equation}
The field equations obtained by varying $e^\gamma{}_\alpha,\,\omega^\gamma{}_\alpha,\,
A^{ik,0},\,A^{ik,2}$ and $\psi^i_\gamma$ are, respectively,
$R^\alpha{}_\gamma=0,\, T^\alpha{}_\gamma=0,\,F^{ki,2}=0,\, F^{ki,0}=0$
and $\mathcal{D}\psi^{i\gamma}=0$, and express the vanishing of the curvatures.
Since $H_2$ in \eqref{HP0} and $I_2$ in \eqref{actionP0} contain all the
gauge fields in the expansion up to order $\lambda^2$, they
determine a CS model based on the Lie superalgebra
$osp(p|2;\mathbb{R})(N,N-1)$ (recall that $V_2=0$) for $N=2$,
$osp(p|2;\mathbb{R})(2,1)$. The $D$=3 $p$-Poincar\'e supergravity model
defined by these equations coincides with that given in \cite{Ho-Iz-Pa-To:95}
when $q$=0, as will be seen in Sec.~\ref{pqgral} below. Therefore, the
three-dimensional ($p,0$)-Poincar\'e gravity is a CS model based on the
superalgebra given by the expansion $osp(p|2;\mathbb{R})(2,1)$.
A dimensional check, using (\ref{dimsuper}), shows that
dim$\,osp(p|2;\mathbb{R})(2,1)= 2({p(p-1)\over 2}+3)+ 2p=\mathrm{dim}\,s\mathcal{P}(p,0)\,$.

\subsection{The general case: $D$=$3$ $s\mathcal{P}(p,q)$ or $(p,q)$-Poincar\'e
supergravity from an expansion of $osp(p+q|2;\mathbb{R})$}
\label{pqgral}
${}$
\\
The first question to address here is what should be the starting
superalgebra that, by expansion, leads to the general $(p,q)$-Poincar\'e
supergravities. One might think of the direct sum $adS(p,q)$
superalgebra itself as a possible candidate. However, besides not being simple,
for $q=0$ it reduces to $osp(p|2;\mathbb{R})\oplus sp(2)$, which is larger than
$osp(p|2;\mathbb{R})$, the expansion $osp(p|2;\mathbb{R})(2,1)$
of which is behind the $(p,0)$-Poincar\'e supergravity models as
shown in Sec.~\ref{p0sugra}. Additionally, it may be checked by direct
computation that an expansion of the $adS(p,q)$ algebra will
not produce a superalgebra leading to a CS action for
$\mathcal{P}(p,q)$-supergravities. These, however, may still be seen to
be CS theories constructed from a specific expansion of
a {\it simple} superalgebra.

    To show that the superalgebra $s\mathcal{P}(p,q)$ of
the ($p,q$)-Poincar\'e supergravities and their CS actions follow
from an expansion of $osp(p+q|2;\mathbb{R})$, we write first the
$osp(p+q|2;\mathbb{R})$ gauge FDA in a convenient form
using eq.~\eqref{curvatures}, where the $i=1,\dots,p+q$ index is split into
two indices $(i,i')$, $i=1,\dots p\,$ and $i'=1,\dots , q$. In this way,
the $osp(p+q|2;\mathbb{R})$ curvatures read
\begin{eqnarray}
\label{osppq}
  F^{ij} &=& dA^{ij} + A^{ik} \wedge A^{kj} + A^{ik'} \wedge
  A^{k'j}+ i\xi^i_\gamma \wedge \xi^{j\gamma} \nonumber\\
  F^{i'j'} &=& dA^{i'j'} + A^{i'k'} \wedge A^{k'j'} + A^{i'k} \wedge
  A^{kj'}+ i\xi^{i'}_\gamma \wedge \xi^{j'\gamma} \nonumber\\
  F^{ij'} &=& -F^{j'i} = dA^{ij'} + A^{ik} \wedge A^{kj'} + A^{ik'} \wedge
  A^{k'j'}+ i \xi^i_\gamma \wedge \xi^{j'\gamma}\nonumber\\
  \Psi^{j\alpha} &=& d\xi^{j\alpha} + f^\alpha{}_\gamma \wedge
  \xi^{j\gamma} + A^{jk} \wedge \xi^{k\alpha} + A^{jk'} \wedge
  \xi^{k'\alpha} \nonumber\\
  \Psi^{j'\alpha} &=& d\xi^{j'\alpha} + f^\alpha{}_\gamma \wedge
  \xi^{j'\gamma} + A^{j'k} \wedge \xi^{k\alpha} + A^{j'k'} \wedge
  \xi^{k'\alpha}\nonumber\\
  \Omega^\alpha{}_\beta &=& df^\alpha{}_\beta + f^\alpha{}_\gamma
  \wedge ^\gamma{}_\beta + i\xi^{i\alpha} \wedge \xi^i_\beta +
  i\xi^{i'\alpha} \wedge \xi^{i'}_\beta \quad ;
\end{eqnarray}
note the mixed indices terms $A^{ij'},F^{ij'}$.
The equations for the exterior derivatives of the curvatures can
be deduced from \eqref{osppq} and will not be needed here. By
simply setting the above curvatures equal to zero, the MC equations of
$osp(p+q|2;\mathbb{R})$ are recovered in the present language.

The next step is to split the underlying vector space of
$osp(p+q|2;\mathbb{R})$ as
$osp(p+q|2;\mathbb{R})= V_0\oplus V_1\oplus V_2$, where $V_1$
is its odd part, and $V_0$ is a subalgebra. The latter will be the
$so(p)\oplus so(q) \oplus sp(2)$ subalgebra, the basis of
which corresponds to the gauge fields
$\{A^{ij},A^{i'j'}, f^\alpha{}_\beta \}$. The
remaining bosonic subsapce, $V_2$, corresponds to the $p\times q$
gauge fields $\{A^{ij'}=- A^{j'i}\}$. The expansions of the gauge
forms and the curvatures of $osp(p+q|2;\mathbb{R})$, subordinated
to the above splitting and up to order $\lambda^2$, are
\begin{eqnarray}
\label{pqexpang}
 f^\alpha{}_\beta & =&  \omega^\alpha{}_\beta + \lambda^2
e^\alpha{}_\beta + o(\lambda^4)\nonumber \\
A^{ij} &=& A^{ij,0} + \lambda^2 A^{ij,2}+ o(\lambda^4)\nonumber \\
A^{i'j'} &=& A^{i'j',0} + \lambda^2 A^{i'j',2}+ o(\lambda^4)
\nonumber\\
A^{ij'} &=&- A^{j'i} =\lambda^2 A^{ij',2} o(\lambda^4)\nonumber\\
 \xi^{i\alpha}
&=& \lambda \psi^{i\alpha} + o(\lambda^3)\nonumber\\
\xi^{i'\alpha} &=& \lambda \psi^{i'\alpha} + o(\lambda^3) \; ,
\end{eqnarray}
where $\xi^{i\alpha,1}\equiv \psi^{i\alpha}\,,\xi^{i'\alpha,1}\equiv \psi^{i'\alpha}$, and
\begin{eqnarray}
\label{pqexpanc}
 \Omega^\alpha{}_\beta & =&  R^\alpha{}_\beta + \lambda^2
T^\alpha{}_\beta + o(\lambda^4)
\nonumber \\
F^{ij} &=& F^{ij,0} + \lambda^2 F^{ij,2} + o(\lambda^4)
\quad , \quad i,j=1,\dots,p \; ,
\nonumber \\
F^{i'j'} &=& F^{i'j',0} + \lambda^2 F^{i'j',2} + o(\lambda^4)
\quad, \quad i',j'=1,\dots,q \; ,
\nonumber\\
F^{ij'} &=& \lambda^2 F^{ij',2} + o(\lambda^4)
\nonumber\\
\Psi^{i\alpha} &=& \lambda \mathcal{D}\psi^{i\alpha} + o(\lambda^3)
\nonumber\\
\Psi^{i'\alpha} &=& \lambda \mathcal{D}\psi^{i'\alpha} +
o(\lambda^3) \quad ,
\end{eqnarray}
where $\Omega^{\alpha\beta}\equiv T^{\alpha\beta}$; note that
$F^{ij',0}\equiv 0$ since the expansions in the $V_2$ subspace
begin already with $\lambda^2$.

Similarly, the curvatures are related with the gauge fields by
\begin{eqnarray}
\label{pqcurv}
F^{ij,0} &=& dA^{ij,0} + A^{ik,0} \wedge A^{kj,0}
\nonumber\\
F^{i'j',0} &=& dA^{i'j',0} + A^{i'k',0} \wedge A^{k'j',0}
\nonumber\\
F^{ij,2} &=& dA^{ij,2} + A^{ik,0}\wedge A^{kj,2} + A^{ik,2}\wedge A^{kj,0} +
i\psi^i_\gamma \wedge \psi^{j\gamma}
\nonumber\\
F^{i'j',2} &=& dA^{i'j',2} + A^{i'k',0}\wedge A^{k'j',2} + A^{i'k',2}\wedge
A^{k'j',0} + i\psi^{i'}_\gamma \wedge \psi^{j'\gamma}
\nonumber\\
F^{ij',2} &=& dA^{ij',2} + A^{ik,0}\wedge A^{kj',2} + A^{ik',2}\wedge
A^{k'j',0} + i\psi^i_\gamma \wedge \psi^{j'\gamma}
\nonumber\\
\mathcal{D}\psi^{j\alpha} &=& d \psi^{j\alpha} +
\omega^\alpha{}_\gamma \wedge \psi^{j\gamma}
+ A^{jk,0}\wedge \psi^{k\alpha} \equiv
D \psi^{j\alpha} +  A^{jk,0}\wedge \psi^{k\alpha}
\nonumber\\
\mathcal{D}\psi^{j'\alpha} &=& d \psi^{j'\alpha} +
\omega^\alpha{}_\gamma \wedge \psi^{j'\gamma} + A^{j'k',0}\wedge
\psi^{k'\alpha} \equiv D \psi^{j'\alpha} +  A^{j'k',0}\wedge
\psi^{k'\alpha}
\nonumber\\
R^\alpha{}_\beta &=& d \omega^\alpha{}_\beta +
\omega^\alpha{}_\gamma
\wedge \omega^\gamma{}_\beta
\nonumber\\
T^\alpha{}_\beta &=&  de^\alpha{}_\beta + \omega^\alpha{}_\gamma
\wedge e^\gamma{}_\beta + e^\alpha{}_\gamma \wedge \omega^\gamma{}_\beta
+ i\psi^{k\alpha} \wedge \psi^k_\beta
+ i\psi^{k'\alpha} \wedge \psi^{k'}_\beta
\nonumber
\\
& & \equiv De^\alpha{}_\beta + i\psi^{k\alpha} \wedge \psi^k_\beta
+ i\psi^{k'\alpha} \wedge \psi^{k'}_\beta  \quad .
\end{eqnarray}
For zero curvatures (and recalling that we took $N_0=2, N_1=1$
and $N_2=2$), eqs.~\eqref{pqcurv} reproduce the MC equations
of $osp(p+q|2;\mathbb{R})(2,1,2)$, the $N_1=1,
N_0=2=N_2$ expansion of $osp(p+q|2;\mathbb{R})$.

    The various fields appearing in \eqref{pqexpanc} and
\eqref{pqcurv} can be interpreted as in the $(p,0)$ case, but now
there are also additional $SO(q)$  $A^{i'j',0}\,$, $A^{i'j',2}$ gauge fields.
Further, \eqref{pqcurv} contains a set of $p\times q$ one-forms
$A^{ij',2}= -A^{j' i,2}$, which transform under the $SO(p)$ and $SO(q)$
groups acting on the corresponding indices.

    The MC equations of $osp(p+q|2;\mathbb{R})(2,1,2)$,
obtained by taking vanishing curvatures in \eqref{pqcurv},
are not yet those of the superalgebra $s\mathcal{P}(p,q)$ of
the $(p,q)$-Poincar\'e supergravity models we are interested in.
The MC equations for $s\mathcal{P}(p,q)$ follow by setting the
curvatures equal to zero in eq.~(7.7) of ref. \cite{Ho-Iz-Pa-To:95},
which in its notation is
\begin{eqnarray}
\label{sppq}
F^{ij} &=& dA^{ij} + A^{ik} \wedge A^{kj}
\nonumber\\
F^{i'j'} &=& dA^{i'j'} + A^{i'k'} \wedge A^{k'j'}
\nonumber\\
G^{ij} &=& dC^{ij} + A^{ik}\wedge C^{kj} + C^{ik}\wedge A^{kj}
- \psi^i_\gamma \wedge \psi^{j\gamma}
\nonumber\\
G^{i'j'} &=& dC^{i'j'} + A^{i'k'}\wedge C^{k'j'} + C^{i'k'}\wedge
A^{k'j'} + \psi^{i'}_\gamma \wedge \psi^{j'\gamma}
\nonumber\\
\mathcal{D}\psi^{j\alpha} &=& d \psi^{j\alpha} +
\frac{i}{2}\omega^a(\gamma_a)^\alpha{}_\gamma  \wedge \psi^{j\gamma} + A^{jk}\wedge
\psi^{k\alpha} \equiv D \psi^{j\alpha} +  A^{jk}\wedge \psi^{k\alpha}
\nonumber\\
\mathcal{D}\psi^{j'\alpha} &=& d \psi^{j'\alpha} +
\frac{i}{2}\omega^a(\gamma_a)^\alpha{}_\gamma \wedge \psi^{j'\gamma} + A^{j'k'}\wedge
\psi^{k'\alpha} \equiv D \psi^{j'\alpha} +  A^{j'k'}\wedge \psi^{k'\alpha}
\nonumber\\
F^a(\omega) &=& d\omega^a-{1\over 2}\epsilon^a{}_{bc}\,\omega^b\wedge\omega^c
\nonumber \\
T^a &=& de^a-\epsilon^a{}_{bc}\,\omega^b\wedge e^c-\frac{1}{4}\bar{\psi}^i\gamma^a\psi^i
-\frac{1}{4}\bar{\psi}^{i'}\gamma^a\psi^{i'}
\nonumber \\
&& \equiv De^a - \frac{1}{4}\bar{\psi}^i\gamma^a\psi^i
-\frac{1}{4}\bar{\psi}^{i'}\gamma^a\psi^{i'} \; .
\end{eqnarray}

   Reverting these expressions to the present notation {\it i.e.},
making the replacements $A^{ij}\rightarrow A^{ij,0} \,,\,
F^{ij}\rightarrow F^{ij,0}$ (and similarly for the primed indices),
$G^{ij}\rightarrow -F^{ij,2}\,,\,
C^{ij} \rightarrow  -A^{ij,2}$ , $G^{i'j'}\rightarrow F^{i'j',2}\,,\,
C^{i'j'} \rightarrow A^{i'j',2}$  setting
$\omega^\alpha{}_\beta={i\over 2}\omega^a\gamma_a{}^\alpha{}_\beta$
(where $\omega^a\propto \epsilon^{abc}\omega_{bc}$),
$R^\alpha{}_\beta= {i\over 2} F^a\gamma_a{}^\alpha{}_\beta$,
$e^\alpha{}_\beta= -2ie^a \gamma_a{}^\alpha{}_\beta$ and
$T^\alpha{}_\beta=-2i T^a \gamma_a{}^\alpha{}_\beta$ in
terms of the $D$=3 gamma matrices, and identifying
$\bar{\psi}$ with the $\psi$ with the index down
(so that $\bar{\psi}^i \gamma^a \psi^i = \psi^i_\alpha (\gamma^a)^\alpha{}_\beta
\psi^{i\beta}$), we see that eq.~\eqref{sppq} reproduces
eq.~\eqref{pqcurv} but for the fifth line
in $F^{ij',2}$. In fact, the algebra generators
corresponding to the $A^{ij',2}$ fields, missing in \eqref{sppq},
generate a $(p\times q)$-dimensional abelian subalgebra
of $osp(p+q|2;\mathbb{R})(2,1,2)$ which is, in fact, an ideal
(clearly trivial if either $p$ or $q$ are zero, which explains
why it was absent in the ($p,0$) case). Denoting this abelian
ideal by $\mathcal{C}$, it is seen that
$osp(p+q|2;\mathbb{R})(2,1,2)/\mathcal{C}= s\mathcal{P}(p,q)$
{\it i.e.}, $osp(p+q|2;\mathbb{R})(2,1,2)$ is an extension of the $D=3$
($p,q$)-superPoincar\'e algebra $s\mathcal{P}(p,q)$ by $\mathcal{C}$,
which is neither central nor semidirect. It is not semidirect because
$s\mathcal{P}(p,q)$ is not a subalgebra of $osp(p+q|2;\mathbb{R})(2,1,2)$:
the commutator of the algebra generators dual to $\psi^i$ and $\psi^{i'}$
have a component in the subspace corresponding to
$A^{ij',2}$, which is not in $s\mathcal{P}(p,q)$.
It is not central either since the abelian ideal $\mathcal{C}$ does
not belong to the centre of $osp(p+q|2;\mathbb{R})(2,1,2)$.

   Again, we can make an easy dimensional check using \eqref{dimsuper}:
dim$\,osp(p+q|2;\mathbb{R})(2,1,2)/\mathcal{C}=
\left(2({p(p-1) \over 2}+{q(q-1) \over 2}+3)+2(p+q)+pq \right)-pq=
\mathrm{dim}\,s\mathcal{P}(p,q)$ by eq.~\eqref{Ppq}.
In fact, it turns out that the $N=2$ term in the expansion
\eqref{CSexpansion} of the action in \eqref{CSsimcom} for
$osp(p+q|2;\mathbb{R})$ actually selects the Lie superalgebra $s\mathcal{P}(p,q)$
rather than $osp(p+q|2;\mathbb{R})(2,1,2)$. Indeed, the first equation in
\eqref{CSsimcom} leads, by expanding in $\lambda$ and selecting
the $\lambda^2$ term, to
\begin{equation}
\label{H2pq}
  H_2 = 2 R^\alpha{}_\gamma \wedge T^\gamma{}_\alpha
  - 2F^{ik,0} \wedge F^{ki,2} - 2F^{i'k',0} \wedge F^{k'i',2}
  - 2i\mathcal{D} \psi^i_\gamma \wedge  \mathcal{D}\psi^{i\gamma}
  - 2i\mathcal{D} \psi^{i'}_\gamma \wedge  \mathcal{D} \psi^{i' \gamma}
\end{equation}
(the first term containing $F^{ij',2}$, $-F^{ij',2} \wedge F^{i'j,2}$,
is of order $\lambda^4)$. Nevertheless, $osp(p+q|2;\mathbb{R})(2,1,2)$ remains
a symmetry of the action by construction. This may be rephrased by noticing that
the Casimir of $osp(p+q|2;\mathbb{R})$, expanded up to order $\lambda^2$,
is degenerate on $\mathcal{C}$, which is its radical as seen
in eq.~\eqref{H2pq}. Thus, once $\mathcal{C}$ is quotiented out, the
resulting metric is no longer degenerate.

   Therefore, the action of the $(p,q)$-Poincar\'e supergravity is found to be
\begin{equation}
\label{I2pq}
  I_2= 2 \int_{\mathcal{M}^3}( R^\alpha{}_\gamma \wedge e^\gamma{}_\alpha
  - F^{ik,0} \wedge A^{ki,2} -  F^{i'k',0} \wedge A^{k'i',2}
  -i \mathcal{D}\psi^i_\alpha\wedge \psi^{i \alpha}
  -i \mathcal{D}\psi^{i'}_\alpha\wedge \psi^{i' \alpha}\,) \;.
\end{equation}
Note that $C^{ij'}$ is absent from both \eqref{H2pq} and \eqref{I2pq};
of course, for $q$=0, both expressions reduce to
eqs~\eqref{HP0} and \eqref{actionP0} respectively. Again, the field equations are
given by the vanishing of the curvatures, and the resulting models
coincide with those of \cite{Ho-Iz-Pa-To:95} (eqs.~(7.11) there).

Therefore, the expansion procedure applied to $osp(p+q|2;\mathbb{R})$
determines the $s\mathcal{P}(p,q)=osp(p+q|2;\mathbb{R})(2,1,2)/\mathcal{C}$
algebras associated with the $(p,q)$-Poincar\'e supergravities
as well as their CS actions.

\section{Concluding remarks}

We conclude with some remarks on the expansion method
and the gauge structure of (super)gravities in general.
The search for a gauge structure of (super)gravities in
various spacetime dimensions for some underlying group is
an old one. On general odd-dimensional spacetimes, gauge theories with
a CS structure are natural candidates; see \cite{Cham:90},
\cite{Ba-Tro-Za:96,Iz-Ro-Mi-Sa-Pe:09} and \cite{Zan:05} for a review.
In fact, for $D$=3, the CS structure of ${\mathcal N}$=1 supergravity
has been known for quite a long time \cite{PvN:85,Witten:88}.
The expansion method was immediately used for lower dimensional
supergravities \cite{AzIzPiVa:02,AzIzPiVa:04}. It has also
been applied to $D$=5 CS AdS gravity in an attempt to
obtain the Einstein-Hilbert lagrangian \cite{Ed-Ha-Tr-Za:06}
from it, albeit the dynamics turns out to be very different from
that of general relativity.

  Even when there is not a clearly singularized initial (super)group
to start with, the expansion procedure may be useful. It was
already pointed out  in the original CJS  $D$=11
supergravity paper \cite{Cre-Ju-Sch:78} that the $OSp(1|32)$
symmetry might play a relevant role in eleven-dimensional
supergravity. It was shown in \cite{D'Au-Fre:82}
that it is indeed possible to associate a Lie algebra to the
$D$=11 CJS supergravity FDA by considering its three-form field
as a composite of one-form fields, which are identified with the
soft MC forms of a superlagebra; this algebra may be said to trivialize the
original FDA one. The general solution that trivializes the FDA structure of $D$=11
supergravity was found in \cite{Ba-Az-Iz-Pi-Va}. It was shown there
that the underlying gauge group structure of CJS supergravity is described
by the members of a one-parameter family of deformations of
a specific expansion of the $osp(1|32)$ algebra,
namely of $osp(1|32)(2,3,2)$, the expansion itself being
excluded. On its part, the M-theory full superlagebra
({\it i.e.}, including the Lorentz automorphism algebra)
may be seen to be \cite{AzIzPiVa:02,AzIzPiVa:04} the
expansion $osp(1|32)(2,1,2)$; also,  the usefulness of
expansions in obtaining the flat limit for the WZ term
of a superstring in anti-de Sitter space has been shown
in \cite{Hat-Sak:01,Hat-Sak:02}.

For the moment, however, it may be said that
a clear understanding of the symmetry structure of
$D$=11 supergravity and of M-theory remains elusive.
At a more modest level, the analysis of the $D$=3 ($p,q$)-Poincar\'e
supergravities presented here does exhibit the expansion character
of the various $\mathcal{P}(p,q)$ Poincar\'e superalgebras behind
them and provides further examples of the usefulness of the expansion
procedure to construct the corresponding CS actions.
\bigskip
\bigskip

\noindent
{\bf Acknowledgements}.
This work has been partially supported by research grants from the
Spanish Ministry of Science and Innovation (FIS2008-01980,
FIS2005-03989), the Junta de Castilla y Le\'on (VA013C05) and EU
FEDER funds.

\end{document}